\documentclass[a4paper,11pt]{article}
\usepackage{jheppub}
\usepackage{graphicx}
\usepackage{xcolor}
\usepackage{booktabs}
 \usepackage{amsmath}
\usepackage[font={small,it}]{caption}
\usepackage{vmargin}
\usepackage{comment}
\usepackage{float}
 \usepackage{multirow}
 \usepackage{siunitx}  

 \usepackage[table]{xcolor}

 \usepackage{amsmath}
 \usepackage{multirow}
 \usepackage{makecell}
 \usepackage{booktabs}   % opcional
 \usepackage{caption}
 \usepackage{orcidlink}
\setpapersize{A4}
\setmargins{2.5cm}       % margen izquierdo
{1.5cm}                        % margen superior
{16.5cm}                      % anchura del texto
{23.42cm}                    % altura del texto
{10pt}                           % altura de los encabezados
{1cm}                           % espacio entre el texto y los encabezados
{0pt}                             % altura del pie de página
{2cm}                           % espacio entre el texto y el pie de página

\newcommand{\ra}[1]{\renewcommand{\arraystretch}{#1}}
\usepackage{subcaption}

\title{Forecasts for lifetime and fraction of Decaying Dark Matter based on redshift distortions from Euclid and BOSS.}

\author{Javier Juarez Jimenez\orcidlink{0009-0004-8314-7776}$^1$ and Ana Avilez-López\orcidlink{0000-0003-2223-4716}$^{1,2}$}

\emailAdd{javier.juarezji@alumno.buap.mx}
\emailAdd{ana.avilezlopez@correo.buap.mx}

\affiliation{$^1$Facultad de Ciencias F\'{\i}sico Matem\'aticas, Benem\'erita Universidad Aut\'onoma de Puebla, Apdo. Postal 1152, Puebla, Pue., M\'exico, \\
$^2$Centro Internacional de F\'isica Fundamental, Benem\'erita Universidad Aut\'onoma de Puebla, Apdo. Postal 1152, Puebla, Pue., M\'exico.}

\abstract{In this work we forecast constraints on models of decaying dark matter (DCDM) by using redshift-space-distortion (RSD) measurements implemented in a Fisher information matrix. In particular, we focus on the fraction of unstable dark matter, $\alpha_{\mathrm{dcdm}}$ respect to the ordinary CDM component, and the decay rate, $\Gamma_{\mathrm{dcdm}}$ as the key parameters of the model. Fiducial values are derived from a MontePython MCMC analysis. The derivatives of the growth-related observable, $f\sigma_8(z)$ with respect to the parameters are numerically around the fiducial model. For the Fisher analysis, we employ mock data designed for upcoming surveys, particularly Euclid and BOSS, where RSD measurements yield constraints on $f\sigma_8.$ Our results show that when both stable and unstable components are allowed, constrains on the DCDM lifetime remain weak, with $\tau_{\mathrm{dcdm}}>1.18$ Gyr. In the limiting case of fully unstable dark matter ($\alpha_{\mathrm{dcdm}}=1$), the uncertainty improves to $\tau_{\mathrm{dcdm}}>235.89$ Gyr. Our findings highlight the potential of RSD probes in testing and complementing decaying dark matter scenarios.}

\begin{document}

\maketitle

\section{Introduction}
Dark Matter (DM) is essential in the structure formation process, although any particle of dark matter has been directly detected so far, we have several observational evidences of its existence due to the effects of its gravitational influence. Therefore, its nature at fundamental level is unknown until today, leading to the famous \textit{Dark Matter Problem}. Among the favorite candidates, one of the most widely studied at are the so called Weakly Interacting Massive Particles (WIMPS) which typically arise within supersymetric extensions of the standard model of fundamental particles as the lightest supersymetric partners of the standard ones. In the standard cosmological scenario, dark matter particles were coupled to the cosmic plasma via a weak interaction in the early universe, and later they decoupled from it through a non-equilibrium process giving rise to the dark matter relic responsible of shaping the distribution of matter that we observe in the late universe \textbf{\cite{Planck:2018vyg}}. This dark matter relic is typically assumed to behave as a non-collisional perfect fluid dubbed cold dark matter (CDM) and at the fundamental level is assumed to be made of electrically and color-neutral stable particles. Although the assumption of the stability of dark matter is reasonable given its not detection so far at human time scales, at cosmic time scales it can be relaxed \textbf{\cite{hambye2010stabilityparticledarkmatter}}. Dark matter might be unstable as long as its lifetime is comparable to the lifetime of the universe without spoiling the non-direct-detection constraints. This motivates the study of decaying cold dark matter (DCDM) scenarios, in which all (or a fraction) of the DM content is unstable. Interestingly, even though the decay rate of dark matter interactions was strongly constrained so that its decaying products remain undetected in earth laboratories, a slow decay of dark matter would alter the fluid properties of dark matter and its gravitational influence and, as a consequence, the process of structure formation would be altered if dark matter is slightly unstable. In recent works bounds on the lifetime of dark matter have been derived according to of cosmological data \cite{Ichiki_2004,Lattanzi_2007,Amigo_2009,Lattanzi:2013uza,Audren:2014bca, Aoyama:2014tga,Poulin_2016,Alvi:2022aam}. Also the thermodynamics of the decaying process has been studied in \cite{Juarez-Jimenez:2025}. An interesting result of the previous work is that the fluid properties of dark matter and its decaying products differ from those species in equilibrium.          

%Aqui me quede 

In this work we aim to determine a forecast for the uncertainties of the lifetime and fraction of DCDM by using the potential of Redshift-space distortions (RSD) as a probe of this models. We focus on two parameters that characterize the $\Lambda$DCDM framework: the decay rate $\Gamma_{\mathrm{dcdm}}$ (or its lifetime $\Gamma^{-1}_{\mathrm{dcdm}} \equiv \tau_\mathrm{dcdm}$) and the present-day fraction of DCDM, $\alpha_{\mathrm{dcdm}}$. Implementing a Markov Chain Monte Carlo (MCMC) analysis with CLASS \cite{lesgourgues2011cosmiclinearanisotropysolving} and MontePython \cite{brinckmann2018montepython3boostedmcmc} we obtained fiducial parameters, and with them, we employed the Fisher matrix formalism \cite{hobson_jaffe_liddle_mukerjee_parkinson_2014, e18060236} to forecast parameters uncertainties. Mock measurements of $f\sigma_8(z)$, which characterize the growth rate of cosmic structures, particularly developed from the Euclid  and BOSS collaboration \cite{Euclid:2021xmh, Hamaus:2020cbu}, are used as observational inputs. By comparing the fiducial DCDM model with this observational data, we can quantify how RSD can contribute to constrain dark matter lifetime and complement existing limits from early and late Universe probes. 

The structure of this paper is as follows: In section 2, we introduce the theoretical framework that describes the $\Lambda$DCDM model and summarize the relevant modifications to the background by introducing linear perturbations. Section 3 describes the MCMC analysis performed with CLASS and MontePython to obtain the fiducial cosmological parameters. In Section 4, we present the Fisher Matrix formalism used to forecast parameter uncertainties. In section 5 we present our results and discussion. Finally, in Section 6 we conclude and describe briefly the implications for future cosmological observations.

\section{\texorpdfstring{$\Lambda$DCDM}{Lambda-DCDM}  cosmological model} 

In this paper, we investigate the simplest scenario of a decaying cold dark matter (DCDM) model, where a cold dark matter \textit{mother} particle decays into a massless \textit{daughter} particle known as dark radiation (DR). This daughter particle belongs to the dark sector, making it a relativistic component that does not interact with the Standard Model particles.

Because dark matter must remains stable on cosmological timescales to fulfill its role in large-scale structure formation and satisfy indirect detection constraints, we focus specifically on particles with lifetimes comparable to the age of the Universe. This approach is further supported by particle physics principles: most Standard Model particles are unstable with lifetimes spanning many orders of magnitude, whereas absolute stability requires imposition of exact symmetries. Since observations do not exclude the possibility of dark matter being unstable yet cosmologically long-lived, this scenario presents a more natural framework than involving unknown symmetry mechanisms.

It can be shown that, after considering a exponential decay of comoving number of DCDM particles and applying the First Law of Thermodynamics \cite{kolb_turner_1994}, that the system of equations governing the energy densities of the dark sector components are: 

\begin{equation}
    \dot{\rho}_{\mathrm{dcdm}}+3 \left(\frac{\dot{a}}{a} \right)\rho_{\mathrm{dcdm}}=-\Gamma_{\mathrm{dcdm}}\rho_{\mathrm{dcdm}}
    \label{density_1}
\end{equation}
\begin{equation}
    \dot{\rho}_{\mathrm{dr}}+4 \left(\frac{\dot{a}}{a} \right)\rho_{\mathrm{dr}}=\Gamma_{\mathrm{dcdm}}\rho_{\mathrm{dcdm}}.
    \label{density_2}
\end{equation}

Here, $\rho_{\mathrm{dcdm}}$ represents the energy density of DCDM, $\rho_{\mathrm{dr}}$ is the DR energy density, and $\Gamma_{\mathrm{dcdm}}$ is the decay rate of DCDM (or $\Gamma^{-1}_{\mathrm{dcdm}} \equiv \tau_{\mathrm{dcdm}}$ its lifetime). The dotted quantities indicate derivatives with respect to proper time. We observe that in absence of decay (i.e, when $\Gamma_{\mathrm{dcdm}}=0$), the energy densities scale as $\rho_{\mathrm{dcdm}}\propto a^{-3}$ and $\rho_{\mathrm{dr}}\propto a^{-4}$, precisely as expected for non-interacting non-relativistic and relativistic particles, respectively. Furthermore, we observe in the right-hand terms of both equations the energy transfer from DCDM particles to DR, which scales proportionally with the mother particle's energy density.

At the perturbative level, the decay of DM introduces a energy and momentum exchange between both dark species \cite{Audren:2014bca}: 

\begin{equation}
    \nabla_{\mu}T^{\mu}_{\mathrm{dcdm}} = -a\Gamma_{\mathrm{dcdm}}\rho_{\mathrm{dcdm}}(1+\delta_{\mathrm{dcdm}}),
\end{equation}

\begin{equation}
    \nabla_{\mu}T^{\mu}_{\mathrm{dr}} = a\Gamma_{\mathrm{dcdm}}\rho_{\mathrm{dcdm}}(1+\delta_{\mathrm{dcdm}}).
\end{equation}

Given that the total energy-momentum remains conserved, we can obtain the following equations for the DCDM and DR fluids in the comoving-Newtonian gauge: 

\begin{equation}
    \delta_{\mathrm{dcdm}}'= -\theta_{\mathrm{dcdm}}-3\dot{\phi} - a\Gamma_{\mathrm{dcdm}}\psi,
\end{equation}

\begin{equation}
    \theta_{\mathrm{dcdm}}' = -\frac{a'}{a}\theta_{\mathrm{dcdm}}+ k^2\psi,
\end{equation}

\begin{equation}
    \delta_{\mathrm{dr}}'= -\frac{4}{3}(\theta_{\mathrm{dcdm}}-3\dot{\phi})+a\Gamma_{\mathrm{dcdm}}\frac{\rho_{\mathrm{dcdm}}}{\rho_{\mathrm{dr}}}(\delta_{\mathrm{dcdm}}-\delta_{\mathrm{dr}}+\psi),
\end{equation}

\begin{equation}
    \theta_{\mathrm{dr}}'=\frac{k^2}{4}\delta_{\mathrm{dr}}-k^2\sigma_{\mathrm{dr}}+k^2\psi-a\Gamma_{\mathrm{dcdm}} \frac{3\rho_{\mathrm{dcdm}}}{4\rho_{\mathrm{dr}}}\bigg (\frac{4}{3}\theta_{\mathrm{dr}}-\theta_{\mathrm{dcdm}}\bigg),
\end{equation}

where $\delta_i$ and $\theta_i$ are the density contrast and velocity divergence of species $i$, respectively, and $\sigma_{\mathrm{dr}}$ denotes the shear of the DR component. 

The decay of dark matter, as seen above, modifies the density and velocity perturbations, leading to a change in the growth of cosmic structures \cite{FrancoAbellan:2021sxk, Enqvist_2015}. In particular, the decay rate $\Gamma_{\mathrm{dcdm}}$ or the fraction of unstable DM, $\alpha_{\mathrm{dcdm}}$ can alter the growth factor $D(z)$ and the growth rate $f(z)=d\ln D/d \ln a$, as well as the combined observable $f\sigma_8$. These effects directly impact redshift-space distortions, making RSD measurements a sensitive probe to constrain the parameters of the $\Lambda$DCDM model.

%%%%%%%%%%%%%%%%%%%%%%%%%%%%%%%
\section{MCMC estimation of \texorpdfstring{$\Lambda$DCDM}{Lambda-DCDM}  parameters}

The effects of the decay of DCDM on the growth rate of structures motivates an exploration of the $\Lambda$DCDM parameter space. In order to establish a fiducial cosmology consistent with current observations, we carried out a Bayesian parameter estimation using the Metropolis-Hashtings algorithm implemented in the public code Monte-Python \textbf{\cite{brinckmann2018montepython3boostedmcmc}} and the CLASS Boltzmann code \textbf{\cite{lesgourgues2011cosmiclinearanisotropysolving}} assuming flat priors on the following set of parameters: 

\begin{equation}
    \{  \omega_b, \theta_s, A_s, n_s, \tau_{\mathrm{reio}},\Omega_{\mathrm{dcdm + dr}},\Omega_{\mathrm{cdm}}, \epsilon_{\mathrm{dcdm}} \}
    \label{param1},
\end{equation}

along with many nuisance parameters not specified here for brevity. The parameters stand for respectively for baryon density today, angular size of sound horizon at decoupling, primordial amplitude of scalar perturbations, spectral index of the initial power spectrum of the scalar perturbations, optical depth in the reonization epoch, density parameter of decaying cold dark matter and its product today, density parameter of stable cold dark matter  and the dcdm decay rate. It may be noticed that $\Omega_{\mathrm{dcdm + dr}}$ is extremely close to $\Omega_{\mathrm{dcdm}}$ if we assume an universe-like dcdm lifetime.

The DCDM model was tested using Planck 2018 collaboration data \textbf{\cite{Planck:2018vyg}} on CMB anisotropies, including: 
\begin{itemize}
    \item  Angular power spectrum measurements of temperature and polarization in the ranges:
    \begin{itemize}
        \item [\textopenbullet] $2 \leq \ell \leq 30$ (low-$\ell$TT)
        \item [\textopenbullet]$30 \leq \ell \leq 2500$ (high-$\ell$TT)
    \end{itemize}
    \item $2 \leq \ell \leq 30$ (low-$\ell$EE).
\end{itemize}

We additionally incorporated late-time baryon acoustic oscillation (BAO) distance measurements from the BOSS collaboration. The results are summarized in Table \ref{Table1} and Figure \ref{fig:planck_bao}.

\begin{table*}\centering
\ra{1.3}
\begin{tabular}{@{}rrrrcrrrcrrr@{}}\toprule
&& Parameter & \phantom{abc}&CMB PLANCK  &\phantom{abc}&CMB PLANCK + BAO BOSS\\
\midrule
&&$100\omega{}_{b }$ &&$2.205^{+0.047}_{-0.042}   $&& $2.238^{+0.020}_{-0.022}$ \\
&&$\Omega_{\mathrm{cdm} }$&&$0.14^{+0.12}_{-0.13}      $ &&$0.232^{+0.046}_{-0.055}$\\
&&$\Omega_{\mathrm{dcdm+dr} }$&&$0.13^{+0.16}_{-0.15}      $&& $0.030^{+0.053}_{-0.043}$\\
&&$\epsilon_{\mathrm{dcdm} }  $ &&$2.3^{+3.4}_{-2.6}         $&& $5.2^{+1.3}_{-1.2}$\\
&&$100\theta{}_{s }$ &&$1.04177^{+0.00094}_{-0.00090}$ &&$1.04188^{+0.00061}_{-0.00063}$ \\
&&$\ln10^{10}A_{s }$ && $3.033^{+0.028}_{-0.033}   $ && $3.050^{+0.046}_{-0.033}$ \\
&&$n_{s }         $ && $0.961^{+0.011}_{-0.010}   $&& $0.9643^{+0.0074}_{-0.0077}$ \\
&&$\tau{}_{\mathrm{reio} } $ &&$0.048^{+0.014}_{-0.017}   $&&$0.058^{+0.021}_{-0.015}$ \\
&&$\Omega{}_{\Lambda }$ &&$0.679^{+0.020}_{-0.021}   $&& $0.688^{+0.011}_{-0.011} $\\
&&$H_0$ &&$66.8^{+1.6}_{-1.5}        $ && $67.52^{+0.76}_{-0.78}     $ \\

\bottomrule
\end{tabular}
\caption{95\% confidence intervals from MCMC chains using Planck 2018 data alone, and Planck 2018 combined with BAO measurements, for a model incorporating both decaying dark matter (DCDM) and stable dark matter components.}
\label{Table1}
\end{table*}

 \begin{figure}
        \centering
        \includegraphics[scale=0.2]{Images/planck_bao.jpg}
        \caption{1D and 2D marginalized probability distributions for a model with both stable and unstable dark matter, using Planck-only data (red) and Planck + BAO (blue). Contours represent 68\% and 95\% confidence levels.
        }
        \label{fig:planck_bao}
    \end{figure}

We find that incorporating late-universe BAO data substantially strengthens the parameter constraints compared to Planck data alone. The best-fit model is dominated by stable dark matter, with decaying dark matter representing only a minor component. This implies either an enhanced decay rate or a shorter lifetime, as the limited decay products produce weaker observable effects, allowing for greater decay activity. Notably, while the marginalized 1D/2D distributions display negative values for both the decaying dark matter density and decay rate, the actual MCMC chains contain no negative samples—a feature we attribute to smoothing artifacts in the posterior visualization. All physical solutions remain within the positive parameter space.
%%%%%%%%%%%%%%%%%%%%%%%%%%%%%%%
\section{Redshift Space Distortions within the \texorpdfstring{$\Lambda$DCDM}{Lambda-DCDM} Model}

In the next years (and decades) large amounts of data of the distribution of matter 
at large scales are to be extracted. These large datasets of information promise to improve our knowledge about the properties of dark matter as a macroscopic system.
In specific, in nearly future surveys such as eBOSS, Euclid and DESI it is planned to measure with high precision the redshift-space-distortions (RSD) produced by peculiar motions of large scale structures. 

Redshift-space distortions (RSD) provide a powerful test of cosmological perturbations \cite{Kaiser:1987qv}, as the perturbation solutions directly determine the observed distortion patterns. These measurable distortions effectively constrain the solutions to the perturbation equations.

 In this work we  are specially interested in determining in which extent future surveys are capable to improve the existing constraints on the lifetime of dark matter. Particularly, our goal is to obtain a forecast for the uncertainties of the lifetime and fraction of unstable cold dark matter whose decaying products correspond to ultrarrelativistic particles by using the Fisher matrix technique we consider mock data from simulations made for future releases of SDSS (BOSS) and Euclid, where RSD were measured in voids-galaxy correlation functions \cite{Euclid:2021xmh,SDSS:2005sxd}.

In an inhomogeneous universe, matter density fluctuations generate gravitational fields, which in turn produce accelerations that act back on the matter itself, creating a peculiar velocity field $\vec{v}_p$ which are deviations from the homogeneous expansion described by Hubble’s law. The total redshift $z_{tot}$ of an object is then the sum of The background cosmological redshift $\bar{z}$ and A small perturbation $\delta z$ due to local inhomogeneities. This leads to a generalized Hubble relation of the form:

\begin{equation}
    z_{tot} = Hr+ \vec{v}_p \cdot \hat{n}. \label{redshift_total}
\end{equation}

In the last equation $\hat{n}$ denotes the unit vector pointing towards the line of sight, which is the only direction contributing to redshift measurements. This allows to derive an expression for the position of an object in real-space position as: $\vec{r}=r\hat{n}$. We now define a redshift-space coordinate  $s$ in analogy to the real-space distance $r$, such that the position vector in redshift-space becomes $\vec{s} = s\hat{n}$. The total redshift $z_{tot}$ can now be expressed through an effective Hubble law that incorporates both the background Hubble flow and peculiar velocity contributions:

\begin{equation}
    z_{tot} = H(t) s,
\end{equation}

therefore, we can find a relation between $s$ and $r$: 

\begin{equation}
    s = r + \frac{\vec{v}_p \cdot \hat{n}}{H},
\end{equation}

Here, $s$ denotes the observable redshift-space distance (distinct from real-space distance $r$). 

Therefore, the difference between the position of an object in redshift and  real spaces is proportional to peculiar velocities. Therefore, RSD are observational signatures of these perturbations and these measurements provide direct measurements of the growth function.

\subsection{Kaisser Effect}

A key RSD is the Kaiser effect, quantified by the void-galaxy correlation function $\xi(r)$. In redshift-space, this takes the form \cite{Hamaus:2017dwj}:

\begin{equation}
    \xi^s(s) = \xi(r)+\frac{f/b}{3}\bar{\xi}(r)+\frac{f}{b}\mu^2 [\xi(r)-\bar{\xi}(r)], 
\end{equation}

where $\xi(r)$ is the correlation function in the configuration space, i.e. it quantifies the extent of clustering of matter at a comoving distance $r$:

\begin{equation}
    \xi(r) \equiv \langle \delta(\vec{x})\delta(\vec{x}+\vec{r})\rangle = \frac{1}{V} \int d^3 \vec{x}\delta(\vec{x})\delta(\vec{x}-\vec{r}) = \int \frac{d^3 k}{(2\pi)^3} P(k)e^{i\vec{k}\cdot\vec{r}}.
    \end{equation}

$f$ is the growth function (see appendix A) , $b$ denotes the galaxy bias parameter accounting for uncertainties in visible galaxy formation processes, and $\mu = \cos\theta$ encodes the geometric orientation between the line-of-sight direction and the void-galaxy separation vector $\vec{r}$. 

Through a Legendre polynomial expansion of the anisotropic void-galaxy correlation function  (see. Ref. \cite{Hamaus:2020cbu}), the Kaiser effect dominates in the monopole ($\ell=0$) and quadrupole ($\ell=2$) moments, which fully characterize the redshift-space distortions:

\begin{equation}
    \xi^s(\vec{s}) = \xi_0^s(s)+\frac{3\mu-1}{2}\xi_2^s(s),
\end{equation}

where $\xi_0^s(s)$ is the monopole term and $\xi_2^s(s)$ is the quadrupole term. Besides, this redshift-space multipoles satisfy the identity:

\begin{equation}
    \xi_0^s(s)-\bar{\xi_0^s}(s)=\xi_2^s(s)\frac{3+f/b}{2f/b}.
\end{equation}

This relation is important for two reasons:  1) it connects redshift-space multipoles without requiring real-space computations, making them independent of any cosmological model and 2) the multipoles depend only on $f/b$. Consequently, Kaiser-effect RSD serve as a direct tool for testing perturbation growth. Results are typically reported as $f/b$ or $f\sigma_8$.

In this work we test the growth function by means of the latter quantity. 

\begin{figure}
        \centering
        \includegraphics[scale=0.35]{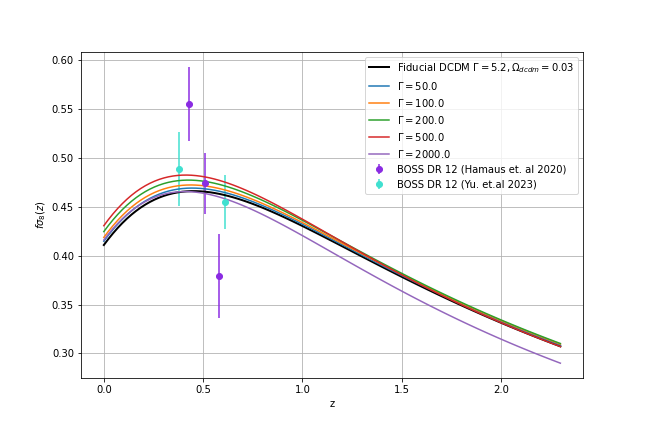}
         \includegraphics[scale=0.35]
        {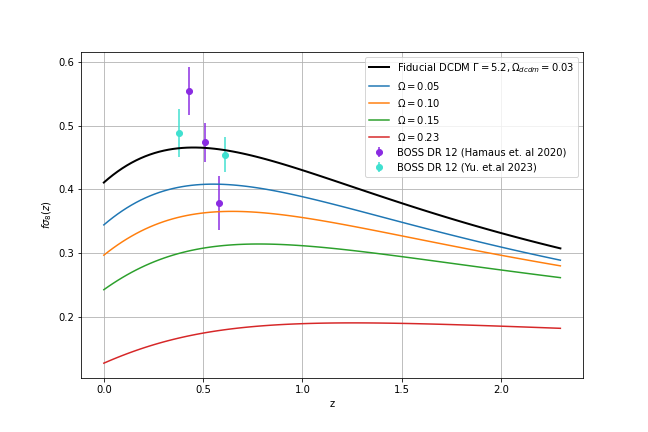}
        \caption{ Predictions of $f\sigma_8(z)$ according to different $DCDM$ models. The black line corresponding to the fiducial model  from Bayesian estimate (gray curve).  The points with error bars correspond to current measurements from different surveys reported in the literature.  }
        \label{fig:fsigma8}
    \end{figure}

\subsection{Alcock-Paczynski Effect}

Another RSD can be measured from geometric deformations in large-scale structures. These occur when an incorrect fiducial cosmology is assumed when converting redshifts to distances, leading to Alcock-Paczynski (AP) distortions \cite{Hamaus:2020cbu}.

When determining the redshift-space separation vector 
$\vec{s}$ between a void-galaxy pair, it is useful to decompose $\vec{s}$ into: 1) the line-of-sight component ($s_{\|}$), derived from the observed redshift separation $\delta z$ and the Hubble parameter as follows: 

\begin{equation}
    s_{\|} = \frac{\delta z}{H(z)},
\end{equation}

and 2) the transverse component ($s_{\bot}$), determined from the angular separation $\delta\vartheta$ and the comoving angular distance $D_A(z)$: 

\begin{equation}
    s_{\bot} = D_{A}(z) \delta\vartheta,
\end{equation}

When computing line-of-sight distances in redshift space, one must adopt a fiducial cosmology for $H(z)$. This introduces rescaling factors between the true (unknown) and fiducial cosmologies:

\begin{equation}
    q_{\bot} = \frac{s_{\bot}^{*}}{s_{\bot}} = \frac{D_{A}^{*}(z)}{D_{A}(z)},
\end{equation}

\begin{equation}
    q_{\|} = \frac{s_{\|}^{*}}{s_{\|}} = \frac{H(z)}{H^{*}(z)},
\end{equation}

where the quantities with  (*)  correspond to the real quantities, and the others correspond to quantities where a fiducial model is assumed.  For systems with average spherical symmetry (e.g., voids or stacked galaxy clusters), we define the AP anisotropy parameter \cite{Euclid:2021xmh}:

\begin{equation}
    \epsilon = \frac{q_{\bot}}{q_{\|}} = \frac{D_{A}^{*}(z) H(z)}{D_{A}(z)H^{*}(z)} \label{epsilon},
\end{equation}

which quantifies the anisotropies due to the AP effect.

\section{Forecast for the uncertainty of the fraction and lifetime of DCDM with RSD from BOSS and EUCLID}
The Fisher matrix is widely employed in cosmology to analyze observational constraints and estimate parameter uncertainties, while also aiding in experiment optimization and forecasting measurement improvements with future data or instruments \cite{HeavensLectures}. The elements of the Fisher matrix are given in terms of the likelihood function $\mathcal{L}$:

\begin{equation}
    F_{\alpha\beta} = \bigg \langle -\frac{\partial^2 \ln \mathcal{L}}{\partial \theta_{\alpha}\partial\theta_{\beta}} \bigg\rangle.
\end{equation}

We have N model parameters labeled by $\theta_i$, and $\alpha,\beta = 1, 2, ..., N$. Therefore, F is a N x N symmetric matrix. The inverse of the Fisher matrix corresponds to the covariance matrix and its diagonal elements are the marginalized constraint to a parameter:

\begin{equation}
    \sigma (\theta_i) = \sqrt{[F^{-1}]_{ii}}.
\end{equation}

Assuming that the uncertainties obey a gaussian distribution, we can write the Fisher matrix elements in a more friendly way:

\begin{equation}
    F_{jk} = \sum_{b} \frac{1}{\sigma_b^2}\frac{\partial O}{\partial \theta_j}\frac{\partial O}{\partial \theta_k}\label{fisher_matrix1},
\end{equation}

where $O$ is an observable quantity which is function of the model parameters and  $\theta_j$ denotes the model parameters.

In order to construct the Fisher matrix for our DCDM model we require to compute the derivatives of the chosen observable with respect to the cosmological parameters of interest. The observable is $\beta(z)\equiv f\sigma_8(z)$. The Fisher matrix is then expressed as:

\begin{equation}
    F_{jk} = \sum_{b} \frac{1}{\sigma_b^2}\frac{\partial \beta}{\partial \theta_j}\frac{\partial \beta}{\partial \theta_k}\label{fisher_matrix2}.
\end{equation}

In our analysis we focus on the DCDM decay rate $\Gamma_{\mathrm{dcdm}}$ (given in units of km s$^{-1}$ Mpc$^{-1}$) and the present-day DCDM+DR density parameter $\Omega_{\mathrm{dcdmdr}}$. We can alternatively use the present-day fraction of DCDM, $\alpha_{\mathrm{dcdm}}$, and can be obtained from the following approximation:

\begin{equation}
    \alpha_{\mathrm{dcdm}} \approx \frac{\Omega_{\mathrm{dcdm+dr}}}{\Omega_{\mathrm{cdm}}},
\end{equation}

which is valid due to our long-lived DCDM hypothesis. We made use of mock catalog data obtained from N-body simulations designed for the planning of future surveys, particularly those of the EUCLID and BOSS collaboration \cite{Euclid:2021xmh, Hamaus:2020cbu}. In these works, cosmic voids are first identified within the matter distribution at different epochs in the simulations, and the multipoles of the void–galaxy correlation function are then estimated in redshift space. From these measurements, redshift-space distortions are quantified, allowing the extraction of $\beta$ at different redshifts along with its uncertainties, which are the ones we use in Eq. (4.4). The used data with its respective uncertainties are listed in Table \ref{tab:datosfb}. 

\begin{table}
\begin{center}
\begin{tabular}{| c | c | c |}
\hline
$z$ & $\beta(z)$  & $\sigma$  \\ \hline
0.38 &  0.489   & 0.038  \\
0.43 & 0.554  &  0.036  \\
0.51 &   0.474  & 0.031\\
0.58 &  0.379   & 0.043 \\ 
0.61  & 0.455  &  0.028  \\ 
0.99& 0.4531& 0.0060\\
1.14& 0.3777& 0.0059\\
1.33& 0.4310& 0.0064\\
2.20& 0.3666& 0.0066\\\hline
\end{tabular}
\caption{
Forecasts on RSD parameter $\beta\equiv f\sigma_8$ from voids in the EUCLID and BOSS mock catalog along with its 1-$\sigma$ uncertainties  \cite{Euclid:2021xmh}.}
\label{tab:datosfb}
\end{center}
\end{table}

As a first step, the compare the fiducial prediction for the $\Lambda$DCDM model for the growth observable $\beta(z)$ with available measurements from the literature. This provides a direct check of the viability of the fiducial cosmology and sets the scale of the statistical uncertainties considered in our Fisher analysis (see Fig \ref{fig:fsigma8}).

Since analytics derivatives of $\beta(z)$ with respect to this parameters are not available, we estimated the derivatives as follows:

\begin{equation}
    \frac{\partial \beta}{\partial\alpha_{\mathrm{dcdm}}}=\frac{\beta(\alpha_{\mathrm{fid}})-\beta(\alpha_{\mathrm{dcdm}})}{\alpha_{\mathrm{fid}}-\alpha_{\mathrm{dcdm}}},
\end{equation}

\begin{equation}
    \frac{\partial \beta}{\partial\Gamma_{\mathrm{dcdm}}}=\frac{\beta(\Gamma_{\mathrm{fid}})-\beta(\Gamma_{\mathrm{dcdm}})}{\Gamma_{\mathrm{fid}}-\Gamma_{\mathrm{dcdm}}},
\end{equation}

where the first terms in the right side of equations (4.6) and (4.7) were computed by running CLASS with all parameters fixed at their fiducial values. On the other hand, the second terms in the above equations were obtained by shifting the parameters (either $\Gamma_{\mathrm{dcdm}}$ or $\alpha_{\mathrm{ddm}}$) by small steps around its fiducial values. To ensure numerical stability, the derivative is recalculated until convergence is achieved. With this, the Fisher matrix can be computed, and its inversion yields the expected uncertainties on the cosmological parameters, $\Gamma_{\mathrm{dcdm}}$ and $\alpha_{\mathrm{dcdm}}$, under the assumption of Gaussian-distributed likelihoods.

We have evaluated the Fisher information matrix for two scenarios: (i) a mixed dark sector containing both stable and unstable dark matter, and (ii) the limiting case in which all of the dark matter is assumed to be unstable, i.e, fixing $\alpha_{\mathrm{dcdm}}=1$. In both cases, the fiducial cosmological parameters were fixed to the best-fit values obtained from our MCMC chains, while the observable $\beta$ was computed at the redshifts where mock measurements are available.

\section{Results and discussion}
The results are summarized in Table \ref{tab:paramsDCDM}.

\begin{table*}[b]
  \centering
  \setlength{\tabcolsep}{14pt}
  \renewcommand{\arraystretch}{1.8}
  \setlength{\arrayrulewidth}{0.5pt}
  \hspace*{-0.7cm}
  \vspace{4pt}
  \begin{tabular}{|c|cc|cc|}
    \hline

    \multirow{2}{*}{\makecell{\textbf{Model}\\ \\ \textbf{Parameter}}}
      & \multicolumn{2}{c|}{\Large DCDM + CDM}
      & \multicolumn{2}{c|}{\Large DCDM only} \\
    \cline{2-5}
      & \textbf{Fiducial value} & \textbf{$\sigma$} & \textbf{Fiducial value} & \textbf{$\sigma$} \\
    \hline\hline

    $\Omega_{\mathrm{dcdm+dr}}$
      & 0.030 & $1.41 \times 10^{-1} $
      & 0.265 (fixed) & - \\
    \hline
    $\alpha_{\mathrm{dcdm}}$
      &  0.001591 & $4.77 \times 10^{-1}$
      & 1 (fixed) & - \\
    \hline
    $\Gamma_{\mathrm{dcdm}}\,[\mathrm{km\,s^{-1}\,Mpc^{-1}}]$
      & 5.2 & $<8.29\times 10^2$
      & 1.8 & $<4.16$  \\
    \hline
   $\tau_{\mathrm{dcdm}}\,[\mathrm{Gyr}]$
      & - & $> 1.18$
      & - & $>234.89 $\\
    \hline
    
  \end{tabular}
  \caption{Forecasts for the cosmological parameters of the models considered in our analysis based on RSD measurements of $\beta = f\sigma_8$. The last line show the uncertainties on the derived parameter $\tau_{\mathrm{dcdm}}=\Gamma_{\mathrm{dcdm}}^{-1} $which represents the DCDM lifetime.}
  \label{tab:paramsDCDM}
\end{table*}

In the mixed scenario, with both stable and unstable components the uncertainties on $\alpha_{\mathrm{dcdm}}$ and $\Gamma_{\mathrm{dcdm}}$  are particularly weak. For instance, the bound of the DCDM lifetime is $\tau_{\mathrm{dcdm}}>1.78$ Gyr, which is significantly below the age of the Universe and therefore fails to exclude scenarios in which a large fraction of dark matter would have already decayed well before the formation of cosmic structures. In contrast, when assuming that the entire dark matter content is unstable ($\alpha_{\mathrm{dcdm}}=1$), we found a lower bound on the DCDM lifetime of $\tau_{\mathrm{dcdm}}>234.89$ Gyr, which is consistent with other works in this area. Based on both results, we can infer a very strong correlation between $\alpha_{\mathrm{dcdm}}$ and $\Gamma_{\mathrm{dcdm}}$. Such a high correlation of both parameters results in situations where one of the values can be significantly modified, paying the price that the other parameter must also change significantly. This was verified in Ref. \cite{Alvi:2022aam}, where they explain that for large values of $\alpha_{\mathrm{dcdm}}$, the upper limit of the decay rate is constrained to small values while for small values (like the one considered in the DCDM + CDM model), the limit stops and relaxes and allows dark matter to decay rapidly. This explains why a weaker bound was obtained from measurements of the Kaiser effect in mock catalogs for EUCLID. 

\begin{figure}[ht]
    \centering

    \begin{subfigure}[t]{0.48\textwidth}
        \centering
        \includegraphics[width=\textwidth]{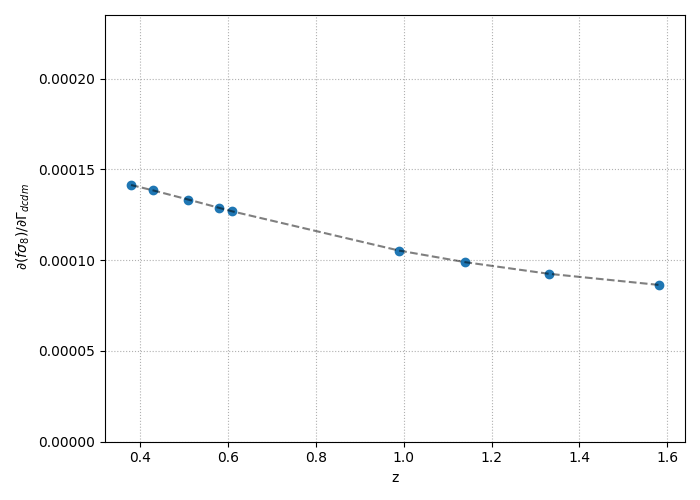}

    \end{subfigure}
    \hfill
    \begin{subfigure}[t]{0.48\textwidth}
        \centering
        \includegraphics[width=\textwidth]{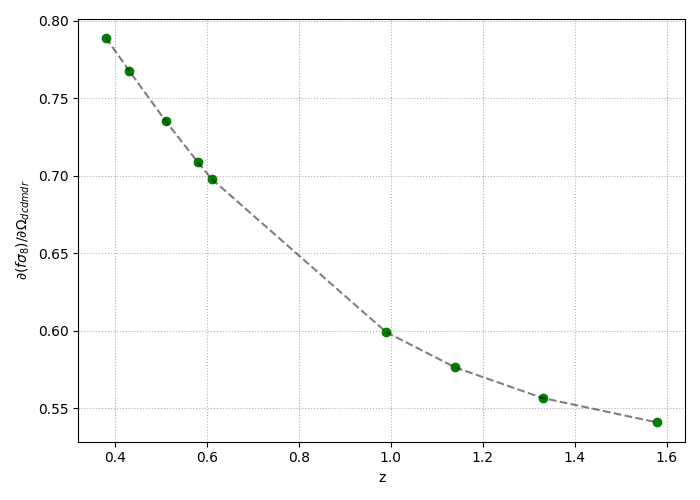}

    \end{subfigure}

    \vskip\baselineskip
    \begin{subfigure}[t]{0.6\textwidth}
        \centering
        \includegraphics[width=\textwidth]{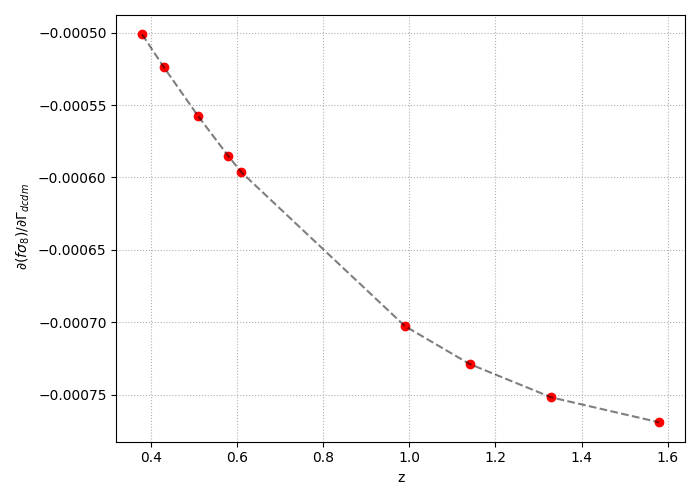}

    \end{subfigure}

    \caption{Numerical derivatives of the observable $\beta(z)=f/b$ with respect to the parameters of the DCDM model. The top panels correspond to the mixed scenario with both stable and unstable dark matter, while the bottom panel shows the fully unstable case ($\alpha_{\rm dcdm}=1$).}
    \label{fig:derivatives}
\end{figure}

Figure \ref{fig:derivatives} displays the numerical derivatives of the observable $\beta(z)$ with respect to the parameters $\Omega_{\mathrm{dcdmdr}}$ and $\Gamma_{\mathrm{dcdm}}$. The first two figures correspond to the DCDM + CDM scenario, and the bottom one correspond to the only DCDM ($\alpha_{\mathrm{dcdm}}=1$) case. In the former, the derivative with respect to $\Gamma_{\mathrm{dcdm}}$ is small in magnitude and exhibits only a mild redshift dependence, remaining nearly constant across the range $
1<z<1.6$. In contrast, when we allow all dark matter to decay, the sensitivity of the observable to $\Gamma_{\mathrm{dcdm}}$ is noticeably larger than in the mixed scenario, and the derivative exhibits a clear monotonic trend with redshift. This behavior reflects the removal of the degeneracy between  $\Gamma_{\mathrm{dcdm}}$ and  $\Omega_{\mathrm{dcdm+dr}}$, since the latter is no longer a free parameter in the fully unstable scenario. As a result, the Fisher analysis provides a stricter bounds on the decay rate and, consequently, the DCDM lifetime.

\section{Conclusions}

In this work we have investigated cosmological models with decaying cold dark matter. Using fiducial cosmologies obtained from MCMC analysis and the Fisher information matrix formalism, we obtained constrains to the model parameters. 

We have shown that in the mixed dark matter scenario, the uncertainties of the parameters translate into a lower bound on the dark matter lifetime of $\tau_{\mathrm{dcdm}}>1.18$ Gyr. If we fixed $\alpha_{\mathrm{dcdm}}=1$, i.e., we only allow all dark matter to be unstable, we expect a stricter constraint of $\tau_{\mathrm{dcdm}}>234.89$ Gyr.

This results illustrate the potential of redshift-space distortion measurements to provide complementary information on dark matter stability. In particular, they highlight the potential of future surveys to improve constraints on decaying dark matter models. 

By combining RSD measurements with other complementary probes, or including larger sets of mocks, like upcoming surveys such as DESI \cite{DESI:2024jxi} (which at the time of writing this paper are not publicly available), DCDM lifetime could potentially be constrained, providing an even more powerful test of the instability of the dark matter particle. 

\appendix
\section{Appendix A: Growth of Density and Velocity Perturbations}

In the linear regime of cosmological perturbation theory, the density contrast modes of matter can be written as \cite{schneider_2016}: 

\begin{equation}
    \delta(a) = D(a) T(k) \delta_0.
\end{equation}

Where $D(a)$ corresponds to the growth function of these matter perturbations and describes how overdensities evolve during the process of structure formation. $\delta_0(k)$ is prescribed by inflation and remains unchanged in the linear regime. Any modification to the scale dependence of matter perturbation is introduced by means of the transfer function $T(k)$. In a similar way, the growth of velocity perturbations  is described by the growth factor \cite{Huterer:2013xky}: 

$$
f(a) = \frac{d\ln D}{d\ln a}.
$$
%%%%%%%%%%%%%%%%%%%%%%
\section{The variance of matter fluctuations}
A commonly used quantity in cosmology is the root mean square of matter fluctuations, where its square $\sigma_{R}^2$ (i.e., the variance of matter fluctuations) is given by \cite{Huterer:2013xky}:

\begin{equation}
    \sigma_{R}^2 = \frac{1}{2\pi^2}\int_{0}^{\infty}k^3P(k)W^2(kR)d\ln k,
\end{equation}

where:

\begin{equation}
    W(x) = \frac{3j_1(x)}{x},
\end{equation}

where $j_1$ is the first-order spherical Bessel function. The quantity $\sigma_R$ quantifies the amplitude of matter density fluctuations averaged over a sphere of radius $R$ at redshift $z$, under the assumption of linear fluctuations. Observations of galaxies within $R = 8h^{-1}~\mathrm{Mpc}$ spheres reveal that their fluctuation amplitude is $\sigma_{8g}^{2} \approx 1$, motivating the standard normalization of present-day matter fluctuations through the parameter $\sigma_8 \equiv \sigma_{8h^{-1}\mathrm{Mpc}}$. The redshift evolution of $\sigma_8$ is commonly used to test the growth function $D(a)$.

\bibliographystyle{ieeetr}
\bibliography{biblio}

\end{document}